\documentclass{emulateapj}

\usepackage{amsmath}
\usepackage{cases}
\usepackage{txfonts}
\usepackage{tipa}
\usepackage{amsfonts}
\usepackage{amssymb}
\usepackage{graphicx}
\usepackage{multirow}

\shorttitle{Method for Making SDSS $u$-Band Magnitude More Accurate}
\shortauthors{Gu et al.}

\begin{document}

\title{A M\MakeLowercase{onte}-C\MakeLowercase{arlo} Method for Making SDSS $u$-Band Magnitude more accurate}

\author{Jiayin Gu\altaffilmark{1}, Cuihua Du\altaffilmark{2}, Wenbo Zuo\altaffilmark{2}, Yingjie
Jing\altaffilmark{2},  Zhenyu Wu\altaffilmark{3}, Jun
Ma\altaffilmark{3} and Xu Zhou\altaffilmark{3}}
\affil{$^{1}$Department of Physics, Wuhan University of Technology, Wuhan 430000, P. R. China; gujiayin12@mails.ucas.ac.cn  \\
$^{2}$School of Physical Sciences, University of Chinese Academy of Sciences, Beijing 100049, P. R. China; ducuihua@ucas.ac.cn \\
$^{3}$Key Laboratory of Optical Astronomy, National Astronomical Observatories, Chinese Academy of Sciences, Beijing 100012, P. R. China}

\begin{abstract}
\par We develop a new Monte-Carlo-based method to convert the SDSS (Sloan Digital Sky Survey)
$u$-band magnitude to the SCUSS (South Galactic Cap of $u$-band Sky
Survey) $u$-band magnitude. Due to more accuracy of SCUSS $u$-band
measurements, the converted $u$-band magnitude becomes more accurate
comparing with the original SDSS $u$-band magnitude, in particular
at the faint end. The average $u$ (both SDSS and SCUSS) magnitude
error of numerous main-sequence stars with $0.2<g-r<0.8$ increase as
$g$-band magnitude becomes fainter. When $g=19.5$, the average
magnitude error of SDSS $u$ is 0.11. When $g=20.5$, the average SDSS
$u$ error is up to 0.22. However, at this magnitude, the average
magnitude error of SCUSS $u$ is just half as much as that of SDSS
$u$. The SDSS $u$-band magnitudes of main-sequence stars with
$0.2<g-r<0.8$ and $18.5<g<20.5$ are converted, therefore the maximum
average error of converted $u$-band magnitudes is 0.11. The potential application of this
conversion is to derive more accurate photometric
metallicity calibration from SDSS observation, especially for those
distant stars. Thus, we can explore stellar metallicity
distributions either in the Galactic halo or some stream stars.

\end{abstract}

\keywords{stars:fundamental parameters-methods:data analysis-star:statistics}

\section{Introduction}

\par It is an increasing perception that the Galactic halo system comprises
at least two spatially overlapping components with different
kinematics, metallicity and spatial distribution \citep{Carollo07,
Carollo10, An13, An15}. Chemical abundance is the direct
observational ingredient in investigating the dual nature of the
Galactic halo. Since the chemical abundance of stars have strong
effect on the emergent flux, especially at blue end, the natural
endeavor is to recover the metal information from large photometric
surveys such as SDSS \citep[Sloan Digital Sky Survey;][]{York00}.
The advantage of photometric metallicity estimate is that the
metallicity information of large numbers of stars can be obtained.

\par Based on the SDSS $ugriz$ photometry,
\cite{Ivezic08} used polynomial-fitting method from spectroscopic
calibration of de-reddened $u-g$ and $g-r$ colors to derive the
photometric metallicity \citep[see also][]{Peng12}. However, due to
the relatively large error of SDSS $u$-band magnitude, only the
metallicities [Fe/H], of stars brighter than $g=19.5$ are obtained.
Combining the more accurate SCUSS \citep{Zhou16} $u$-band
photometry, SDSS $g$ and $r$ photometry, \cite{Gu15} developed a
three-order polynomial photometric metallicity estimator, in which
$u$-band magnitude can be used to faint magnitude of $g=21$.
However, both estimator developed by \cite{Ivezic08} and \cite{Gu15}
based on polynomial-fitting have their intrinsic drawback that they
can not be extended to metal-poor end. In order to solve this
problem, \cite{Gu16} (hereafter denoted as Paper I) devised a
Monte-Carlo method to estimate stellar metallicity distribution
function (MDF) which appears particularly good at both metal-rich
and metal-poor ends. The natural forward step is to combine the
SCUSS $u$, SDSS $g, r$ photometry with the method introduced in
Paper I to investigate the MDF of the Galactic halo stars. But only
those stars in South Galactic cap are surveyed by SCUSS. How can we
estimate the photometric metallicity distribution of faint stars
(deep in Galactic halo) in both South and North hemisphere? This
paper provides a new method to achieve this goal. Due to the fact that SCUSS
$u$ is more accurate than SDSS $u$, we convert SDSS $u$ to SCUSS $u$
using a Monte-Carlo method, through which we make the converted $u$
magnitude becomes as accurate as SCUSS $u$ magnitude.

\par We organize this paper as follows. In Section 2,
we take a brief overview of the SDSS and SCUSS. The technical
details for converting SDSS $u$ to SCUSS $u$ are presented in
section 3.  Section 4 evaluate the effectiveness of this conversion.
The discussion of the potential application of the conversion is given in Section 5.

\section{SDSS and SCUSS}

\par The SDSS is a digital multi-filter imaging and spectroscopic redshift survey using a dedicated 2.5 m wide-angle
optical telescope at Apache Point Observatory in New Mexico, United States \citep{Gunn06}.
It began operation in 2000, and finally over 35\verb"%" of the sky is covered, with about 500 million photometrically surveyed objects and
more than 3 million spectroscopically surveyed objects. Five bands
($u$, $g$, $r$, $i$, and $z$) are used to simultaneously measure the objects magnitude, respectively
with the effective wavelength of 3551, 4686, 6165, 7481, and
8931 \AA. The limit magnitudes of $u$, $g$, $r$, $i$, and $z$ are 22.0, 22.2, 22.2, 21.3, and 20.5, respectively \citep{Abazajian04}.
The relative photometric calibration accuracy for $u$, $g$, $r$, $i$, and $z$ are
2\%, 1\%, 1\%, 1\% and 1\%, respectively \citep{Padmanabhan08}.
Other technical details about SDSS can be found on the SDSS website
\emph{http://www.sdss3.org/}, which also provide interface for the public data access.

\par
\begin{figure}
\includegraphics[width=1.0\hsize]{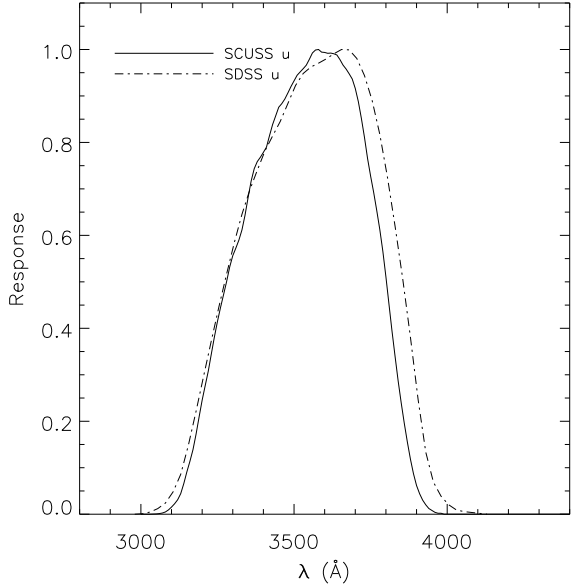}
\caption{Response curves of both the SCUSS $u$ and the SDSS $u$
filters. Atmospheric extinction at the airmass of 1.3 is taken into
account, and both curves are normalized to their maxima. }
\label{Response}
\end{figure}

 The South Galactic Cap $u$-band Sky Survey (SCUSS) is an international cooperative project that is
jointly undertaken by National Astronomical Observatories of China and Steward Observatory of University of Arizona.
It utilizes the 2.3 m Bok telescope located on Kitt Peak to photometrically survey the stars in
the South Galactic Cap in $u$ band with effective wavelength of 3538 \AA.
This project started in the summer of 2009, began its observation in the fall of
2010, completed in the fall of 2013, and finally about 5000
deg$^2$ area ($30^\circ<l<210^\circ,~ -80^\circ<b<-20^\circ$) were surveyed.
It's main goal is to provide the essential input data to the Large Sky
Area Multi-Object Fiber Spectroscopic Telescope (LAMOST) project
\citep{Zhao06}.
Figure \ref{Response} shows the similarity in response curve between the $u$ band filter of SCUSS and that of SDSS.
The limit magnitude for point sources is about 23.2 mag with a 5-minute exposure time,
and is about 1.5 mag deeper than that of SDSS $u$ band magnitude \citep{Jia14, Peng15}.
In Table~\ref{summary}, we provide a brief summary of SCUSS.
The more detailed information and data reduction
about SCUSS can be found in \cite{Zhou16, Zou15, Zou16}, and the SCUSS website \emph{http://batc.bao.ac.cn/Uband/},
which also provides interface for public data access.

\begin{table}
\begin{center}
\caption{Brief summary of SCUSS}
\begin{tabular}{p{3.6cm}p{3.6cm}}
\hline
\hline
Telescope & 2.3 m Bok telescope \\ \hline
Site & Kitt Peak in Arizona \\ \hline
CCD & 2$\times$2 4k$\times$4k CCD array \\ \hline
Exposure time & 300 s \\ \hline
Filter Wavelength & 3538 \AA  \\ \hline
Filter FWHM & 520 \AA \\ \hline
Magnitude Limit & 23.2 mag \\ \hline
Survey Area & $\sim$ 5000 deg$^2$ \\ \hline
Observation Period & 2010$\sim$2013 \\
\hline
\end{tabular}
\label{summary}
\end{center}
\end{table}

\begin{figure}
\includegraphics[width=1.0\hsize]{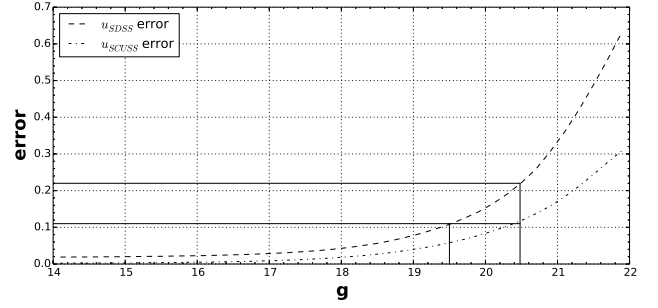}
\caption{Average $u$ (SDSS and SCUSS) error as a function of
$g$-band magnitude. Main-sequence stars with $0.2<g-r<0.8$ are
selected. It is obvious that the error of SDSS $u$ is much larger
than that of SCUSS $u$, especially at the faint end. } \label{Error}
\end{figure}

\par As shown in Figure \ref{Error}, the average error of SCUSS $u$ and SDSS $u$ of numerous main-sequence stars
with $0.2<g-r<0.8$ are plotted as functions of $g$-band magnitude.
It clearly shows that the error of SDSS $u$ is much larger than that
of SCUSS $u$ on the whole, especially at the faint end. The
spectroscopically surveyed stars has limiting magnitude of $g=19.5$.
Coincidentally, the error of SDSS $u$ limits the application of
photometric metallicity estimates in the range of $g<19.5$. From
Figure \ref{Error}, we find that the error of SDSS $u$ is about 0.11
when $g=19.5$. So we set 0.11 as the maximum error. Beneath the
error of 0.11, the SCUSS $u$ corresponds to the range of $g<20.5$.
However, the SDSS $u$ error is up to 0.22 when $g=20.5$. In the
following, we will convert SDSS $u$ to SCUSS $u$ for stars brighter
than $g=20.5$ so that the error of converted $u$ don't exceed 0.11.
Here, we only convert the SDSS $u$ with $18.5<g<20.5$ for
main-sequence stars. Since $g$-, $r$-band magnitudes are much more
accurate than $u$, we assume that they are absolutely precise, at
least in the considered $g$-band magnitude range. So the error of
$u-g$ is the direct consequence of the error of $u$.

\section{Method}

\par For each object surveyed by SCUSS, we can identify the same object
from SDSS catalog by matching their positions. So in the merged catalog, each star has the following
information: position ($ra~\&~dec$), SCUSS $u$-band magnitude and its error, SDSS $u, g, r, i, z$-band magnitudes
and their error and extinction. Here, the extinction for
SDSS $u$-band magnitude is also used by SCUSS $u$-band magnitude.
Throughout this paper, magnitudes and colors are understood that they have been corrected
for extinction and reddening following \cite{Schlegel98}.
We select the stars from SCUSS catalog by the following criteria: \\
1. $18.5<g<20.5$;  \\
2. $0.2<g-r<0.8$;  \\
3. $0.6<(u-g)_{SDSS}<2.2$; \\
4. $0.6<(u-g)_{SCUSS}<2.2$; \\
5. main-sequence stars are selected by only including those objects at distances smaller than
$0.15$ mag from the stellar locus described by the following equation \citep{Juric08}:
\begin{align}
(g-r)=& 1.39\{1-exp[-4.9(r-i)^3-2.45(r-i)^2 \nonumber \\
 & -1.68(r-i)-0.05]  \} \nonumber
\end{align}  \\
6. we further refine the selection of main-sequence stars by only including those objects
at distances smaller than $0.3$ mag from the stellar locus
described by the following equation \citep{Jia14}:
\begin{align}
(u-g)_{SDSS}=exp[-(g-r)^{2}+2.8(g-r)-1]  \nonumber
\end{align} \\

\begin{figure*}
\includegraphics[width=1.0\hsize]{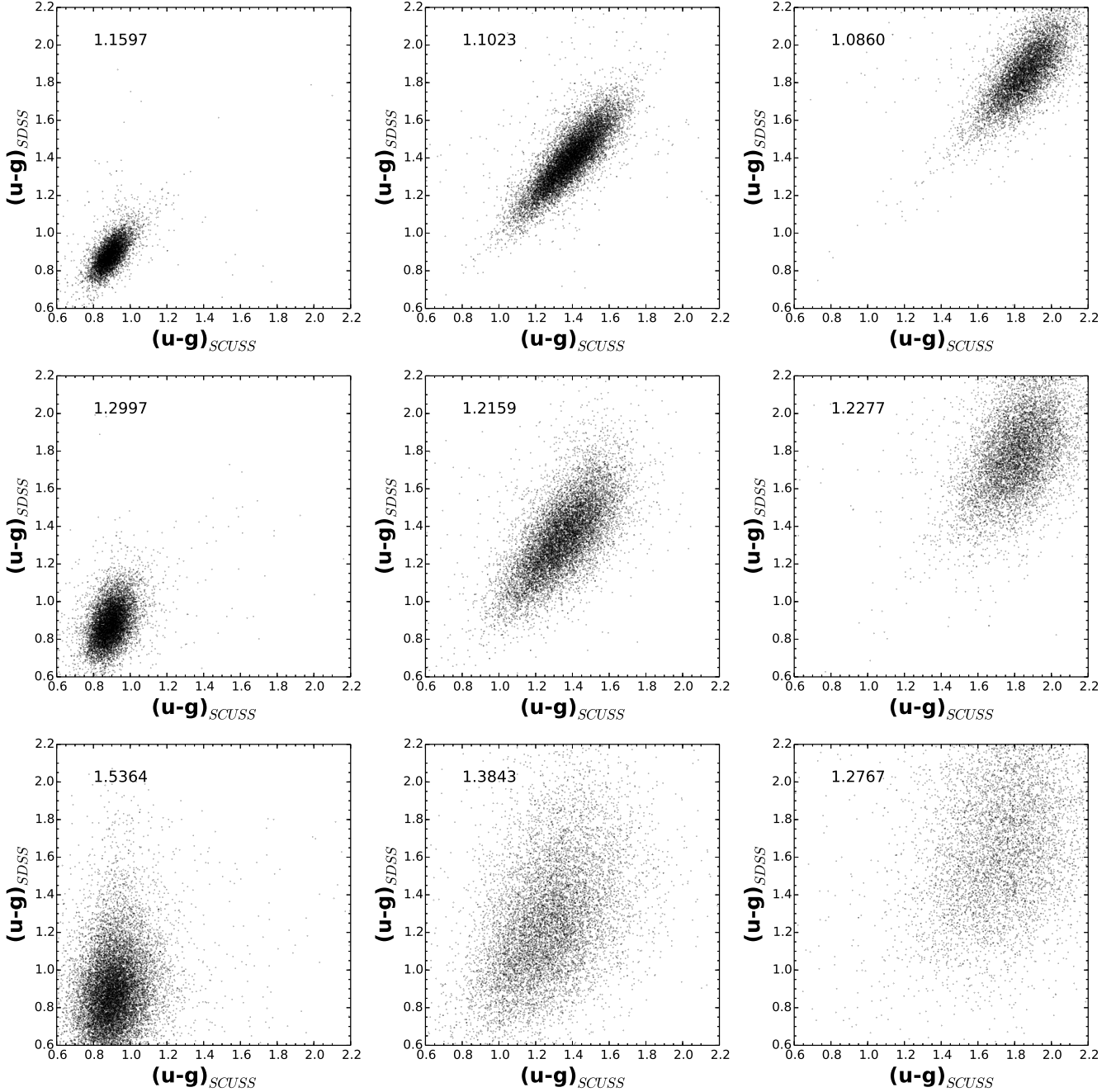}
\caption{Two-color diagrams for $(u-g)_{SCUSS}$ versus
$(u-g)_{SDSS}$. Main-sequence stars in different magnitude and color
range are selected. Stars for panels from top row to bottom row are
with $18.5<g<18.6$, $19.4<g<19.5$ and $20.4<g<20.5$ respectively.
Stars for panels from left column to right column are with
$0.2<g-r<0.3$, $0.5<g-r<0.6$ and $0.7<g-r<0.8$ respectively. The
numbers shown in each panel are the ratios of standard deviation
between $(u-g)_{SDSS}$ and $(u-g)_{SCUSS}$. These numbers are
greater than one, which imply that SCUSS $u$ is more accurate than
SDSS $u$. Additionally, these numbers become larger as $g$-band
magnitude becomes fainter, and the largest one corresponds to bottom
left panel (blue and faint).} \label{Color_scatter}
\end{figure*}

\par We divide the color range of $0.2<g-r<0.8$ into 6 equal bins, and also divide the magnitude range of $18.5<g<20.5$ into 20 bins.
Thus, we totally get 120 $0.1\times0.1$ mag$^2$ bins, and designate each square bin by an $index$ computed in the following manner:
\begin{align}
index=int((u-g-0.2)/0.1)*20+int((g-18.5)/0.1) \nonumber
\end{align}
where the symbol $int$ stands for the integer portion. In this way
the $index$ takes value from 0 to 119. Main-sequence stars whose
colors and magnitudes match a position specified by $index$ will be
used to construct a ``convertor''. Thus, we will totally obtain 120
convertors, and each convertor is denoted as $convertor[index]$. In
the following, each convertor has the form of $16\times16$ array in
which each element is further denoted as $convertor[index][i][j]$,
where $i$, $j$ range from 0 to 15. Each main-sequence star that
is associated with one convertor is further classified with two
labels of integer number, $i$ and $j$, which can be computed in the
following manner:
\begin{align}
i=int(((u-g)_{SDSS}-0.6)/0.1) \nonumber \\
j=int(((u-g)_{SCUSS}-0.6)/0.1) \text{,} \nonumber
\end{align}
where the symbol $int$ also stands for the integer portion.
Each element in each convertor array records the number of stars whose $(u-g)_{SDSS}$ and $(u-g)_{SCUSS}$ colors match its position.
We use $convertor[index][i][:]$ to denote the set of 16 numbers of $convertor[index][i][j]$ for $j$ taking integer values from 0 to 15.
The maximum value of the $convertor[index][i][:]$ is further denoted as $max[index][i]$.

\par Figure \ref{Color_scatter} shows
the two-color diagrams for $(u-g)_{SCUSS}$ versus $(u-g)_{SDSS}$.
Main-sequence stars in different magnitude and color range are
selected.  Stars for panels from top row to bottom row are with
$18.5<g<18.6$, $19.4<g<19.5$ and $20.4<g<20.5$, respectively.  Stars
for panels from left column to right column are with $0.2<g-r<0.3$,
$0.5<g-r<0.6$ and $0.7<g-r<0.8$, respectively. Each one corresponds
to one convertor array. The more scattered the points in each
panel, the larger error $u$-band magnitude it implies.  As shown
in Figure \ref{Color_scatter}, the error of $(u-g)_{SDSS}$ is larger
than that of $(u-g)_{SCUSS}$, especially for those faint stars. The
comparison of error for each panel is quantized by the ratio of
standard deviation between $(u-g)_{SDSS}$ and $(u-g)_{SCUSS}$ that
is shown. For the bottom left panel of the diagram, the ratio has
the maximum value of 1.536 when comparing with others. This panel
corresponds to the fainter and bluer stars that are reasonably belong
to the Galactic halo.

\par The central idea for converting SDSS $u$ to SCUSS $u$ is that we obtain the color distribution of converted $u-g$
according to the inputting distribution of $(u-g)_{SDSS}$ and the
scatter diagram in each panel of Figure \ref{Color_scatter}. The
scatter diagram of $(u-g)_{SCUSS}$ versus $(u-g)_{SDSS}$ is now
explained as the consequence of probability. More points in a small
region imply that a star have higher probability to locate in it. We
reproduce a given distribution by the Monte-Carlo method. The
converted $u-g$ should be considered as same as $(u-g)_{SCUSS}$.
Here, in order to distinguish the converted $u-g$ from original
$(u-g)_{SDSS}$ and $(u-g)_{SCUSS}$, we denoted converted $u-g$ with
a subscript, namely $(u-g)_{CONV}$.

\par For $n$ stars corresponding to $index=a$, $i=b$, we generate $n$ random numbers
according to the distribution exhibited by the 15 numbers from
$convertor[a][b][0]$ to $convertor[a][b][15]$. The obtained $n$
random numbers are all real numbers from 0 to 15. Then, the $n$
random numbers are converted to $(u-g)_{CONV}$ values by
$(u-g)_{CONV}=0.1*r+0.6$, where $r$ is one random number. How to
generate a sequence of random numbers that just comply with a given
distribution? It is explained as follows. Suppose that there are two
stochastic variables, $X$ and $Y$, which can be assigned
a random number generating function $X=rand_{X}()$ and $Y=rand_{Y}()$, respectively.
In each trial, we obtain a random
number pair ($X$, $Y$), where $X$ is modulated to take the
uniform-probability distributed random real number from 0 to 15. For
any star, whose $index$ and $i$ has determined, $Y$ is modulated to
take the uniform-probability distributed random real number from 0
to $max[index][i]$. When $Y\le convertor[index][i][int(X)]$
($int(X)$, the integer portion of $X$), we record $X$ as a useful
value, and otherwise discard it. By numerous trials, we obtain a
sequence of random numbers $\{X_{1}, X_{2}, X_{3}, \cdots\}$ that
follow the same probability distribution as those recorded in
$convertor[index][i][:]$. Here, because $convertor[index][i][j]$ can
be equal to zero for some $j$ values, we can discard them and record
the non-zero elements and their positions in a new array. Through
this method, the sampling efficiency can be improved greatly.

\section{Testing}

\begin{figure}
\includegraphics[width=1.0\hsize]{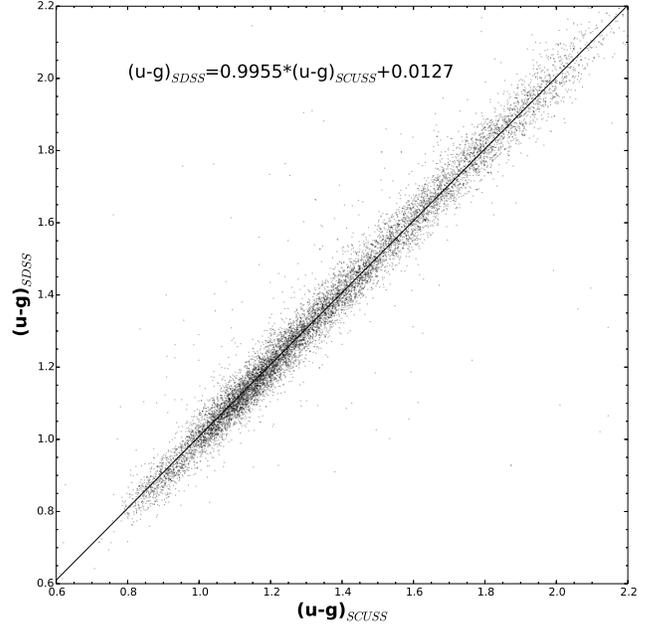}
\caption{Two-color diagrams for $(u-g)_{SCUSS}$ versus $(u-g)_{SDSS}$. Main-sequence stars with $0.2<g-r<0.8$ and $16.99<g<17.01$ are selected.
The data are fitted with a linear line, with the expression shown in the figure.
The slope almost equal to 1.}
\label{Color_scatter_fitting}
\end{figure}

\par From the top three two-color diagrams of Figure \ref{Color_scatter},
we find that $(u-g)_{SDSS}$ may be expressed as a linear function of
$(u-g)_{SCUSS}$. For the selected stars with $18.5<g<18.6$, the error of
$u$ plays the minor role for the distribution of points in
these diagrams.  If the $u$-band (both SDSS and SCUSS) magnitudes
were absolutely precise , the resulting transformation relation is
supposed as follows:

\begin{align}
(u-g)_{SDSS}=k*(u-g)_{SCUSS}+h \text{,} \nonumber
\end{align}
where $k$ is the slope and $h$ represents a constant.

In evaluating which color (either $(u-g)_{SDSS}$ or $(u-g)_{SCUSS}$)
has greater error,  the reliability of the standard deviation ratio
shown in Figure \ref{Color_scatter} depends on the assumption of
$k\approx1$. In Figure \ref{Color_scatter_fitting}, we plot a
two-color diagram of $(u-g)_{SCUSS}$ versus $(u-g)_{SDSS}$ for
main-sequence stars with $0.2<g-r<0.8$ and $16.99<g<17.01$. We also
notice that the error of $u$-band magnitude at the bright magnitude
$g=17$ is small, and therefore its effect on the color distribution
in Figure \ref{Color_scatter_fitting} can be neglected. The trend of
$(u-g)_{SDSS}$ versus $(u-g)_{SCUSS}$ is fitted by a line,with the
expression shown in the figure. The slope $k=0.9955$ is almost equal
to 1. The assumption of $k\approx1$ holds on. We can evaluate which
$u$ has greater error by dispersion degree of points in Figure
\ref{Color_scatter}. In addition, for convenience we may also
approximately assume that SDSS $u$ and SCUSS $u$ are from the same
photometric system, since they are almost similar if neglecting the
error, as shown in Figure \ref{Color_scatter_fitting}.

\begin{figure*}
\includegraphics[width=1.0\hsize]{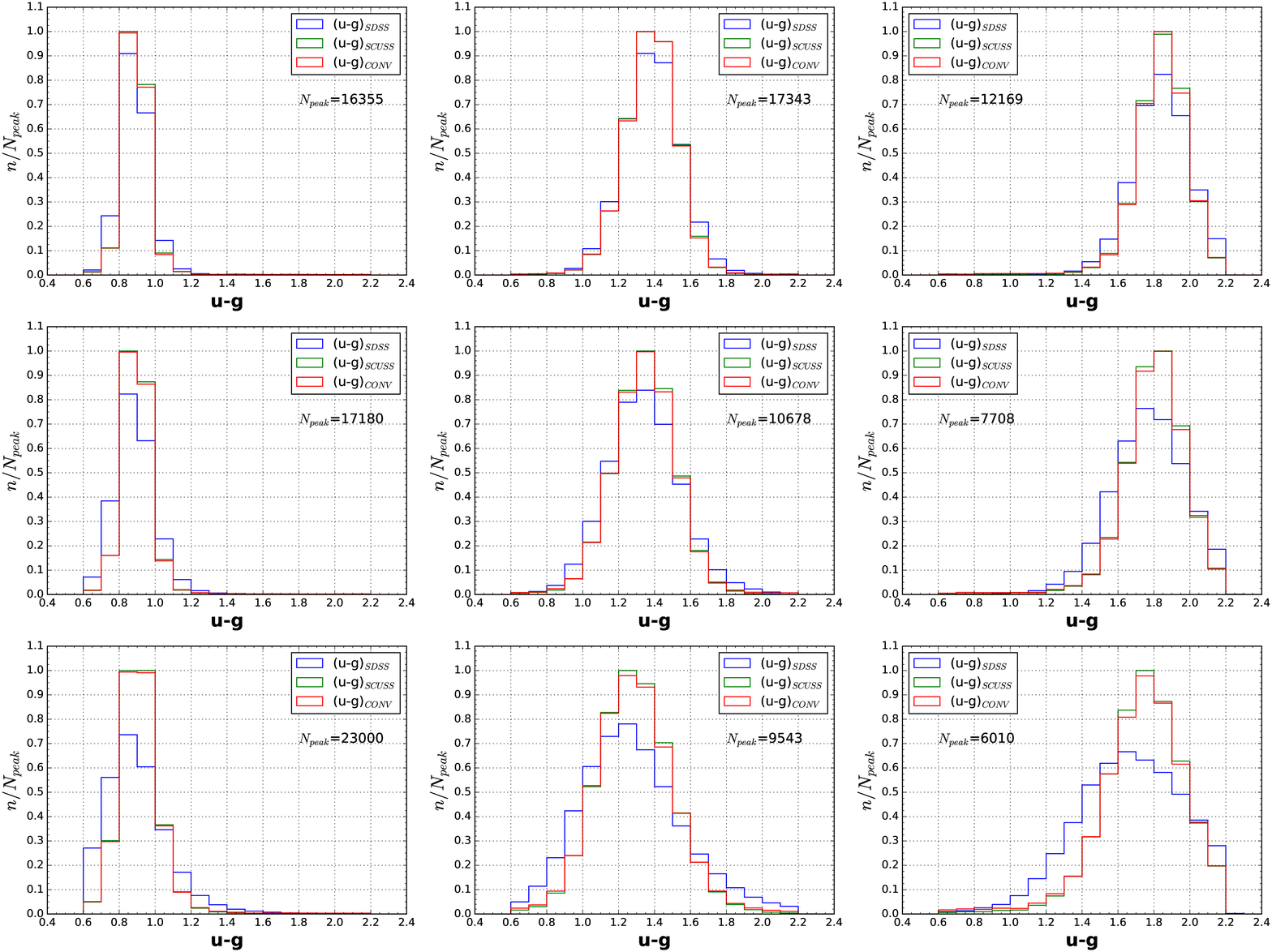}
\caption{Histograms of distribution of $(u-g)_{SDSS}$ (blue),
$(u-g)_{SCUSS}$ (green) and $(u-g)_{CONV}$ (red) with different
magnitude and color range. Stars for panels from top row to bottom
row are with $18.5<g<19$, $19.3<g<19.7$ and $20<g<20.5$. Stars for
panels from left column to right column are with $0.2<g-r<0.3$,
$0.5<g-r<0.6$ and $0.7<g-r<0.8$, respectively. The histograms in
each panel are normalized to the maximum, with actual peak values
labeled.  The histograms of $(u-g)_{CONV}$ and histograms of
$(u-g)_{SCUSS}$ nearly coincide, directly reflecting the
effectiveness of the conversion from SDSS $u$ to SCUSS $u$.}
\label{Color_histogram}
\end{figure*}

\par In order to evaluate the effect of this conversion,
 we plot the histograms of distribution of $(u-g)_{SDSS}$, $(u-g)_{SCUSS}$ and $(u-g)_{CONV}$ for main-sequence stars
with different magnitude and color ranges in Figure
\ref{Color_histogram}. The top three panels show color distribution
of stars with $18.5<g<19$, the middle three with $19.3<g<19.7$, and
the bottom three with $20<g<20.5$. Corresponding with the color
range, stars for panels from left column to right column are with
$0.2<g-r<0.3$, $0.5<g-r<0.6$ and $0.7<g-r<0.8$, respectively. The
histograms in each panel are normalized to the maximum, with the
actual peak values labeled. It is clear that the profiles of the
histograms of $(u-g)_{CONV}$ in each panel are almost same as those
of $(u-g)_{SCUSS}$. This effect indicates that the conversion has
the ability to make the error of $u_{SDSS}$
smaller, as small as that of $u_{SCUSS}$. Actually, the distribution
of $(u-g)_{CONV}$ will completely coincide with that of
$(u-g)_{SCUSS}$ as long as the number of stars selected is large
enough for the histogram. After all the convertor arrays are
constructed by the data of SCUSS $u$. Thus, for larger sky area in
which there have no SCUSS $u$,  the convertor array could be used to
make the error of SDSS $u$ smaller.  As a result, the error of the
converted $u$ magnitude when $g=20.5$ is equal to the original error
of SDSS $u$ when $g=19.5$. However, we are still cautious that
to what extent this conversion method can diminish the $u$ magnitude error cannot
be fully tested until a deeper survey is available.

\section{Discussion}

\par As we all know, $u$-band measurements is very important to
derive the photometric metallicity and therefore to construct a
precise MDF.  Because of the relatively shallow survey limit
($u\sim22$) and the relatively large error in the SDSS $u$-band near
the faint end, the application of the photometric metallicity
estimates is greatly restricted in the range of $g<19.5$, an
insufficient depth to explore the distant halo and substructures.
However, the SCUSS $u$ is 1.5 mag deeper than SDSS $u$, and its
error is smaller than SDSS $u$ error on the whole. The potential
application of the conversion from SDSS $u$ to SCUSS $u$ is very
important to derive relative accurate photometric metallicities of
distant stars. In Paper I, we developed a new method to estimate the
photometric metallicity distribution of large number of stars.
Compared with other photometric calibration methods, this method in
Paper I effectively reduces the error induced by the method itself,
and therefore enables a more reliable determination of the
photometric MDF. However, another error source still matters: the
error of SDSS $u$-band magnitude. This error behavior
 limits the application of the method in the range of
$g<19.5$ in Paper I. This range is same as that of Ivezi\'c et al.'s (2008)
photometric metallicity estimator. The more accurate SCUSS $u$-band
measurements guarantee the accuracy of the stellar distribution in
$u-g$ versus $g-r$ panel, and it extends the application of method
in Paper I to even fainter stars. Thus, the photometric MDF of
distant stars such as halo stars or some stream stars can be
estimated.

However, only the stars in South Galactic Cap are surveyed by SCUSS
which have relatively more accurate $u$ band magnitude, how to
derive the photometric metallicity of stars in the North Galactic
hemisphere? The conversion from SDSS $u$ to SCUSS $u$ statistically
diminish the error of $u$-band magnitude, which make it possible to
estimate the photometric MDF of stars in the whole sky. In this
study, we have done the conversion for stars in $18.5<g<20.5$. The
conversion combined with the method introduced in Paper I enable us
to estimate the photometric metallicity distribution function for
stars at least in the range of $g<20.5$, which is 1 mag deeper than
that of spectroscopically-surveyed stars. So we can study the
chemical structure of the Galactic halo more detailed. Besides the
application described above, the more accurate $u$ band magnitude
from the conversion, can be applied to address other scientific
issues.

\section*{ACKNOWLEDGMENTS}

\par This work was supported by joint fund of Astronomy of
the National Natural Science Foundation of China and the Chinese
Academy of Science, under Grants U1231113.  This work was also by
supported by the Special funds of cooperation between the Institute
and the University of the Chinese Academy of Sciences. In addition,
this work was supported by the National Natural Foundation of China
(NSFC, No.11373033, No.11373035), and by the National Basic Research
Program of China (973 Program) (No. 2014CB845702, No.2014CB845704,
No.2013CB834902).

\par We would like to thank all those who participated in observations and
data reduction of SCUSS for their hard work and kind cooperation.
The SCUSS is funded by the Main Direction Program of Knowledge
Innovation of Chinese Academy of Sciences (No. KJCX2-EW-T06). It is
also an international cooperative project between National
Astronomical Observatories, Chinese Academy of Sciences and Steward
Observatory, University of Arizona, USA. Technical supports and
observational assistances of the Bok telescope are provided by
Steward Observatory. The project is managed by the National
Astronomical Observatory of China and Shanghai Astronomical
Observatory.

\par Funding for SDSS-III has been provided by the Alfred P. Sloan Foundation, the Participating Institutions, the National Science Foundation, and the U.S. Department of Energy Office of Science. The SDSS-III web site is \emph{http://www.sdss3.org/}.

\par SDSS-III is managed by the Astrophysical Research Consortium for the Participating Institutions of the SDSS-III Collaboration including the University of Arizona, the Brazilian Participation Group, Brookhaven National Laboratory, Carnegie Mellon University, University of Florida, the French Participation Group, the German Participation Group, Harvard University, the Instituto de Astrofisica de Canarias, the Michigan State/Notre Dame/JINA Participation Group, Johns Hopkins University, Lawrence Berkeley National Laboratory, Max Planck Institute for Astrophysics, Max Planck Institute for Extraterrestrial Physics, New Mexico State University, New York University, Ohio State University, Pennsylvania State University, University of Portsmouth, Princeton University, the Spanish Participation Group, University of Tokyo, University of Utah, Vanderbilt University, University of Virginia, University of Washington, and Yale University.

\end{document}